\documentclass[journal]{IEEEtran}

\usepackage[utf8]{inputenc}  
\usepackage[T1]{fontenc}

\usepackage{caption}
\usepackage{graphicx}
\usepackage{biblatex}  
\usepackage{multirow}
\usepackage{textcomp,booktabs}
\usepackage[usenames,dvipsnames]{color}
\usepackage{colortbl}
\usepackage{makecell}
\usepackage{amssymb}
\definecolor{mygray}{gray}{.9}

\ifCLASSINFOpdf
  
\else
  
\fi

\begin{document}

\title{Applications of Multi-Agent Deep \\Reinforcement Learning with Communication \\in  Network Management: A Survey}

\author{
Yue Pi\IEEEauthorrefmark{2}\IEEEauthorrefmark{3}, \IEEEauthorblockN{Wang Zhang\IEEEauthorrefmark{2},  Yong Zhang\IEEEauthorrefmark{4}, Hairong Huang\IEEEauthorrefmark{4}, Baoquan Rao\IEEEauthorrefmark{4}, \\ \footnotemark{\textsuperscript{1}}Yulong Ding\IEEEauthorrefmark{2}, \footnotemark{\textsuperscript{1}}Shuanghua Yang\IEEEauthorrefmark{2}\IEEEauthorrefmark{5}} 

\IEEEauthorblockA{\IEEEauthorrefmark{2}Shenzhen Key Laboratory of Safety and Security for Next Generation of Industrial Internet and Department of Computer Science and Engineering, Southern University of Science and Technology, Shenzhen, China} 
\IEEEauthorblockA{\IEEEauthorrefmark{3}Peng Cheng Laboratory, Shenzhen, China}
\IEEEauthorblockA{\IEEEauthorrefmark{4}Huawei Technologies Co., Ltd.} 
\IEEEauthorblockA{\IEEEauthorrefmark{5}Department of Computer Science, University of Reading, UK}

}
\maketitle
\footnotetext[1]{Corresponding author.}

\markboth{}%
{Shell \MakeLowercase{\textit{et al.}}: Applications of Multi-Agent Deep Reinforcement Learning Communication in Network Management: A Survey}

\maketitle

\begin{abstract}
With the advancement of artificial intelligence technology, the automation of network management, also known as Autonomous Driving Networks (ADN), is gaining widespread attention. The network management has shifted from traditional homogeneity and centralization to heterogeneity and decentralization. Multi-agent deep reinforcement learning (MADRL) allows agents to make decisions based on local observations independently. This approach is in line with the needs of automation and has garnered significant attention from academia and industry. In a distributed environment, information interaction between agents can effectively address the non-stationarity problem of multiple agents and promote cooperation. Therefore, in this survey, we first examined the application of MADRL in network management, including specific application fields such as traffic engineering, wireless network access, power control, and network security. Then, we conducted a detailed analysis of communication behavior between agents, including communication schemes, communication content construction, communication object selection, message processing, and communication constraints. Finally, we discussed the open issues and future research directions of agent communication in MADRL for future network management and ADN applications.

\end{abstract}


\IEEEpeerreviewmaketitle

\section{Introduction}

\IEEEPARstart{T}{he} improvement of general computing capabilities has led to the continuous expansion of various production networks. The traditional centralized network management mode has gradually become unable to meet the demand. Therefore, deploying distributed and decentralized network engineering technologies to improve network performance under different network resource and service demand conditions is an essential trend in future network management. Network management involves monitoring and controlling network resources to ensure effective network operation. This comprises several technologies catering to different scenarios and requirements. Traditional network management technologies usually rely on heuristic methods, but with the increase of new devices and services, the network is becoming more extensive and more complex, and it is difficult to model and predict network devices accurately. At the same time, traditional methods are hard to meet the demand for automation and intelligent management for future network management. 

In recent years, due to the development of artificial intelligence technology, technologies such as Machine Learning, Deep Learning (DL), and Reinforcement Learning (RL) have started to be implemented in network management. Multi-Agent Deep Reinforcement Learning (MADRL) based on DL and RL is considered an effective technique to provide AI network solutions for critical problems in the future Internet\cite{li2022applications}. In MADRL, each network entity is regarded as an agent, gathers dynamic and uncertain environmental information, and independently takes action. However, in some partially observable distributed multi-agent systems, due to the mutual influence of adaptation strategies, the system is vulnerable to non-stationary problems \cite{papoudakis2019dealing}. Sharing observations, intentions, or experiences helps agents understand the environment and achieve stable learning. Meanwhile, effective communication between agents can be crucial in promoting cooperation among them \cite{zaiem2019learning}. 

The current communication protocols between agents are usually predefined, resulting in a considerable increase in the total communication overhead as the number of agents increases. An alternative method called emergent communication enables agents to autonomously learn communication protocols and select the communication objects through interaction. However, it also faces several challenges, such as unreadable emergent messages by humans and requiring massive training before deployment. Moreover, its current application in network management is relatively limited; therefore, the reliability and feasibility of emergent communication need further validation.

This survey summarizes the communication behavior of multi-agent systems in network management. We first introduce and summarize the MARL system with agent communication currently applied in the four research directions of network management. Then we conduct a detailed analysis of the communication schemes of the system, the type of communication messages, the selection of communication objects, the message processing method of communication, and the constraints in agents' communication. Finally, we propose existing issues and suggest future research directions for the communication of agents in network management.

\begin{table*}[tp]
\normalsize
\centering
\caption*{TABLE I}
\caption*{\scshape Existing Surveys on MADRL Communication or Network Management}
\label{table1}
\begin{tabular}{|c|c|c|c|} 
\hline
\multirow{3}{*}{\textbf{Work}} & \multicolumn{2}{|c|}{\textbf{MADRL}} & \multirow{3}{*}{\textbf{Scope}} \\ \cline{2-3}
 
 &  \makecell[c]{Network \\ Management} & Communication & \\ \hline

\cite{li2022applications} & \checkmark &  & Future Internet \\ \hline

\cite{feriani2021single} & \checkmark &  & B5G/6G wireless network \\ \hline

\cite{xiao2021leveraging} & \checkmark &  & Traffic Engineering \\ \hline

\cite{althamary2019survey} & \checkmark &  & Vehicular Networks \\ \hline

\cite{zhu2022survey} &  & \checkmark & Agent Communication \\ \hline

\cite{chafii2023emergent} &  & \checkmark & Emergent Communication \\ \hline

\makecell[c]{Our \\ Survey} & \checkmark & \checkmark & \makecell[c]{Network Management + Agent \\ Communication} \\

\hline
\end{tabular}
\label{table_MAP}
\end{table*}


\begin{figure*}[tp]
  \centering
  \includegraphics[width=1\linewidth]{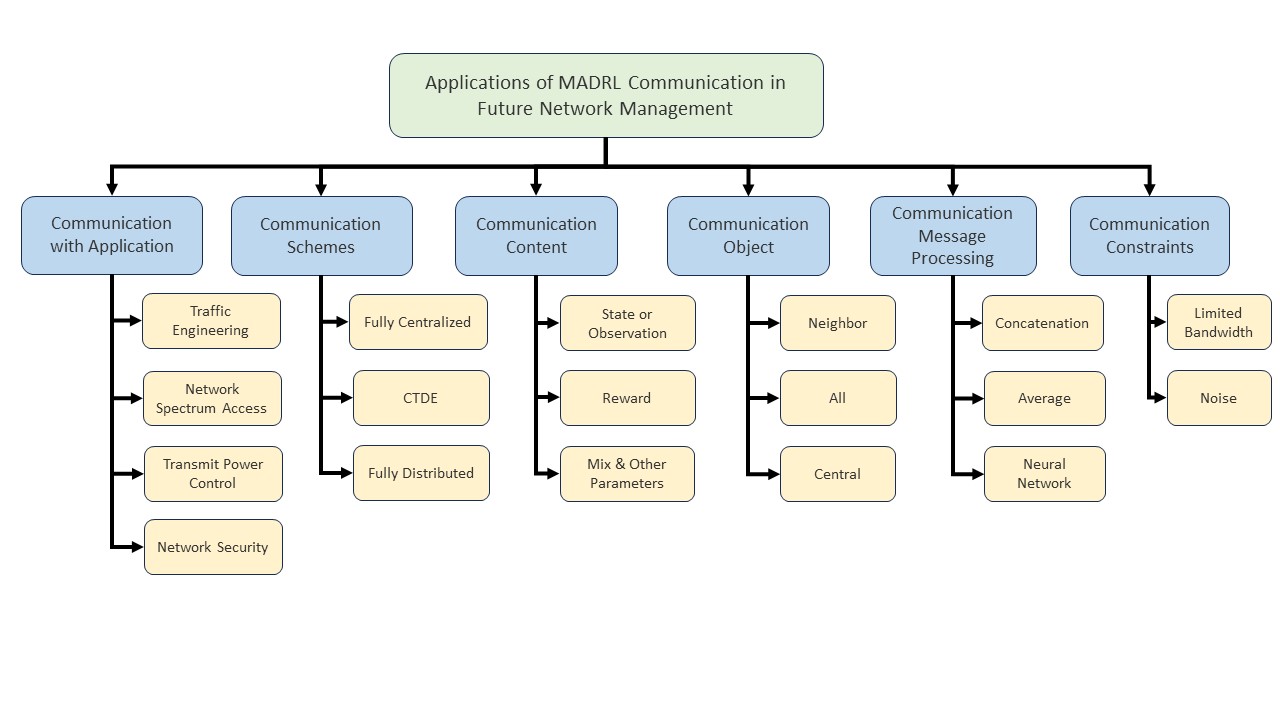}
  \caption{A classification of the applications of MADRL in network management with communication.}
  \label{fig:1}
\end{figure*}

\subsection{Related Existing Survey and Our Scope}
Since MADRL has been widely used to solve critical problems in network management, many surveys have summarized and classified relevant literature from different aspects, as shown in Table I. For example, The authors in \cite{li2022applications} conducted a detailed analysis of the application of MARL in five fields in the future network: network access, edge computing, routing, unmanned aerial vehicle (UAV), and network security. In \cite{feriani2021single}, the authors focus on the MADRL application in the future 6G network management. Furthermore, some works focus on specific scenarios, such as the authors in \cite{xiao2021leveraging} detailing DRL and MADRL's application in Traffic Engineering. In \cite{althamary2019survey}, the authors analyzed some applications of MARL in vehicular networks. In addition to network management, we also pay attention to the reviews of communication between multiple agents. For example, in \cite{zhu2022survey}, the authors analyzed the characteristics of communication between agents in MADRL from nine dimensions in detail. The authors in \cite{chafii2023emergent} discuss the potential application of Emergent Communication in future wireless networks. 

However, to the best of our knowledge, there is currently no comprehensive review of communication in MADRL systems applied to network management. Therefore, this survey focuses on reviewing, analyzing, and comparing existing MADRL works with communication among agents applied to network management, as shown in Figure 1. We classify and summarize six key characteristics of agent communication and conduct a separate analysis of each of them. Meanwhile, we introduce several MADRL emergent communication algorithms that have the potential for application to network management in the future. 

\subsection{Contributions and Survey Organization}
Our contributions are as follows:

\begin{itemize}
\item[$\bullet$] We introduce the application of MADRL in network management and classify the properties of the agent communication in the MADRL system. We summarize the settings for agent communication that could be selected in different application scenarios and show the potential of using these algorithms to solve networking issues in future network management.

\item[$\bullet$] We summarize several uncommon characteristics in MADRL related to communication in network management and discuss the unresolved issues and future research directions of using MADRL systems with agent communication in future network management.

\end{itemize}

The rest of this paper is organized as follows:

Section II summarizes the recent works for the MADRL with communication in the four research directions of network management, including traffic engineering, network spectrum access, transmit power control, and network security. Section III classifies these works by their communication schemes. In Section IV, we introduce the classification of the message content of agent communication. Sections V and VI introduce the selection scheme of communication objects and message processing methods of MADRL's work with communication. Section VII provides a summary of the communication constraints of agent communication. In Section VIII, we discuss the open issues and future research directions of agent communication of the applications of MARL for network management. The conclusions are given in Section IX.

The summary of the classification of MADRL communication in network management is shown in Table II.

\begin{table*}[tp]
\centering
\caption*{TABLE II}
\caption*{\scshape Summary of MADRL Communication in Network Management}
\label{table2}
\begin{tabular}{|c|c|c|c|c|c|c|}
\hline
\textbf{Work} & \textbf{\makecell[c]{Application\\Background}} & \textbf{\makecell[c]{Communication\\Schemes}} & \textbf{\makecell[c]{Communication \\ Content}} & \textbf{\makecell[c]{Communication\\object}} & \textbf{\makecell[c]{Message\\Processing}} & \textbf{\makecell[c]{Communication\\Constraints}}\\
\hline
[31] & \multirow{11}{*}{\makecell[c]{Traffic \\ Engineering}} & Fully Distributed & State & Neighbor & Concatenation & -\\ \cline{1-1} \cline{3-7}

[34] &  & CTDE & State & Neighbor & Concatenation & Limited Bandwidth\\ \cline{1-1} \cline{3-7}

[55] &   & Fully Distributed & Reward & All & - & -\\ \cline{1-1} \cline{3-7}

[56] &   & Fully Distributed & State & All & - & -\\ \cline{1-1} \cline{3-7}

[57] &   & Fully Distributed & State & All & - & Noise\\ \cline{1-1} \cline{3-7}

[35] &   & Fully Distributed & Reward & Neighbor & - & Noise\\ \cline{1-1} \cline{3-7}

[32] &   & Fully Distributed & M\&OP & Neighbor & Neural Network & -\\ \cline{1-1} \cline{3-7}

[36] &   & Fully Distributed & Reward & All & - & -\\ \cline{1-1} \cline{3-7}

[58] &   & Fully Distributed & Reward & All & - & -\\ \cline{1-1} \cline{3-7}

[37] &   & CTDE & State & Neighbor & Concatenation & -\\ \cline{1-1} \cline{3-7}

[33] &   & CTDE & State & Neighbor & Neural Network & -\\
\hline

[40] & \multirow{7}{*}{\makecell[c]{Network \\ Spectrum \\ Access}} & CTDE & State & All & Concatenation & Noise\\ \cline{1-1} \cline{3-7}

[39] &   & Fully Centralized & M\&OP & Central & Neural Network & Noise\\ \cline{1-1} \cline{3-7}

[54] &   & Fully Centralized & M\&OP & Central & - & -\\ \cline{1-1} \cline{3-7}

[42] &   & Fully Distributed & State & Neighbor & Concatenation & -\\ \cline{1-1} \cline{3-7}

[38] &   & Fully Distributed & M\&OP & All & Concatenation & -\\ \cline{1-1} \cline{3-7}

[43] &   & Fully Distributed & M\&OP & All & Average & -\\ \cline{1-1} \cline{3-7}

[41] &   & CTDE & State & Neighbor & Neural Network & Noise\\ 
\hline

[44] & \multirow{7}{*}{\makecell[c]{Transmit \\ Power \\ Control}} & CTDE & State & All & Concatenation & -\\ \cline{1-1} \cline{3-7}

[49] &   & Fully Distributed & State & Neighbor & - & -\\ \cline{1-1} \cline{3-7}

[45] &   & CTDE& State & Neighbor & - & -\\ \cline{1-1} \cline{3-7}

[46] &   & Fully Distributed & State & Neighbor & Concatenation & -\\ \cline{1-1} \cline{3-7}

[50] &   & Fully Distributed & State & Neighbor & Concatenation & -\\ \cline{1-1} \cline{3-7}

[47] &   & Fully Distributed & State & All & Concatenation & -\\ \cline{1-1} \cline{3-7}

[48] &   & Fully Distributed & Reward & All & Concatenation & -\\ 
\hline

[51] & \multirow{3}{*}{\makecell[c]{Network \\ Security}} & CTDE & Reward & All & Concatenation & -\\ \cline{1-1} \cline{3-7}

[52] &   & CTDE & Reward & All & Concatenation & -\\ \cline{1-1} \cline{3-7}

[53] &   & Fully Centralized & M\&OP & Central & - & Limited Bandwidth\\ 
\hline

\end{tabular}
\label{table_MAP}
\end{table*}

\section{MADRL Communication with Application}

The information interaction between agents can help each agent in the system better understand the variation of the global environment and the impact of other agents on the environment, thereby improving the system's overall performance. Moreover, agents can effectively achieve better cooperation by sharing information such as observations, actions, or reward values, thereby achieving better team performance \cite{feriani2021single}. In wireless communication environments, where communication could be unreliable or costly, some works assume that agents don't communicate with each other. For example, wireless heterogeneous networks \cite{li2019multi,zhao2019deep,zhang2020deepNA,meng2020power,guo2020joint}, cognitive radio (CR) networks \cite{kaur2020energy,yang2020partially}, wireless cellular networks \cite{zhang2020deep,doshi2021deep,li2019multi33,yang2022dynamic,li2022dynamic}, UAV networks\cite{zhang2020multi,zhang2020uav,qiu2022data,shamsoshoara2019distributed,cui2019multi}, vehicular networks \cite{liang2019spectrum,nguyen2019distributed}, and IoT \cite{sharma2019multi,salh2022intelligent}. 

In this survey, we only focus on the existing applications of MADRL in network management with agent communication. This section categorizes works with agent communication by their application: traffic engineering, network spectrum access, transmit power control, and network security.

\subsection{Traffic Engineering}
Traffic engineering refers to optimizing routing paths for traffic flows based on their characteristics and balancing the load between different switches, routers, and links in the network. This requires the system to achieve dynamic and real-time monitoring, analysis, control, and prediction of traffic status on the network \cite{xiao2021leveraging}. Traditional traffic engineering methods or routing protocols are defined by fixed rules \cite{shah2021routing}, which make it challenging to achieve the autonomous monitoring and dynamic network management requirements in the network. Nevertheless, MADRL methods enable each network entity, as an agent, to independently learn an adaptive routing strategy and dynamically change its routing rules. Furthermore, the MADRL system with agent communication enables agents to make dynamic decisions that consider the current network status and the impact of other agent routing policies. Therefore, MADRL, especially with agents' communication, might be an effective solution for traffic engineering in future network management.

The authors of \cite{zhao2020improving} analyze and verify the improvement of agents' communication on Autonomous Systems (ASs) throughput. They propose a MADRL routing method for ASs, where each AS on the Internet acts as an agent. ASs select the best next-hop AS for different flows to maximize system throughput, according to the observation of oneself and adjacent agents. The authors test the system performance when the agent communicates with neighbor agents within different ranges. The simulation results show that the system with agent communication performs better than the system where agents only use their local observations to make discussions. Moreover, improving the communication ranges results in an increase in the system's average throughput. The author points out that interacting with more agents can help agents better understand the network state and reduce the impact of non-stationary issues in the distributed MADRL system.

In \cite{alliche2022impact}, the authors design two communication mechanisms for the packet routing problem: value sharing and model sharing. Agents establish communication connections through signaling and sharing their Deep Neural Networks (DNN) models (model sharing) or exchange estimates of end-to-end delay (value sharing) with neighboring agents. Regarding packet loss ratio and average end-to-end packet delay, model sharing is always better than value sharing when replay-memory sizes are large enough. 

In \cite{jiang2018graph}, graph convolution reinforcement learning is used to adapt to the dynamics of the underlying network graph in multi-agent environments. Unlike the traditional MADRL approach, it regards each data packet in the network rather than the router as an agent. The packet as the agent is a node in the graph, and the local observation encoding of the agent is a node feature. Agents formulate cooperation policies by using convolutional neural networks (CNNs) to learn the feature information of nodes from the communication between agents. The experiment shows that compared to the routing algorithm using the Deep Q network, the routing algorithm that learns to encode and communicate messages through Graph Neural Networks (GNN) performs better in system throughput and latency.

In addition, several works allow information exchange among agents but do not discuss the impact of communication between agents on the system. For example, In \cite{you2020toward}, the authors propose Deep Q-routing, based on the Deep Q network, to reduce the average transmission time of packets in Autonomous Systems. The network model is represented as a directed graph. The routers, which are the nodes on the network graph, act as agents. Routers choose one of the neighboring agents as the next hop node for the data packet based on their local observations and neighbors' information.

The study presented in \cite{chen2021universal} has introduced a traffic allocation plan for the Crowd-sourced Licast services (CLS) system. The authors use the Augmented Graph Model to model a large-scale CLS system as a multi-hop routing problem. As an agent, the CLS system's network node determines the flow path in the network depending on observation and the other agents' local rewards. 

The authors of \cite{modi2023multi} found that designing a single reward function can lead to agents becoming lazy or selfish. Thus, the authors designed a delay-tolerant global and local reward function. By capturing information from neighboring agents during training and implicitly sharing packet TTL during execution, a higher level of cooperation between agents is achieved in the SDN environment.

Work \cite{zhang2021sac} proposes a MADRL routing algorithm in a mixed-distributed IoT system. Agents collect and exchange self-observation information containing queue length and remaining energy with other agents to maximize the total amount of data transmission for long-term goals while reducing device energy consumption.

\subsection{Network Spectrum Access}
The network spectrum access method aims to improve spectrum efficiency by allocating spectrum resources reasonably and avoiding quality of service (QoS) reduction caused by competition among network entities for spectrum resources. The algorithm based on MARL can enable network entities to adaptively select spectrum resources based on environmental changes, and is therefore widely used. Agents in the MARL system typically make decisions to improve system spectrum utilization and throughput by collecting and sharing information such as channel state information (CSI), QoS, etc. The information exchange between agents can help them perceive the network status and make better channel allocation policies. 

The authors of \cite{dong2022dynamic} argue that the different types of agent information exchange can impact the MADRL dynamic spectrum access system. They explore a MADRL algorithm that maximizes the information rate of secondary users while ensuring that the primary user's spectrum usage is not affected. Moreover, the performance of three agent communication models is tested: (1) agents only exchanging reward information, (2) agents exchanging reward and state, and (3) agents exchanging reward, state, and action. The result of the experiment shows that as the types of message parameters exchanged between agents increase, the system can adapt more quickly in the spectrum environment and avoid more channel conflicts. Moreover, more information exchange can also improve the sum information rate of the secondary users.

The authors in \cite{wang2020learn} propose a communication information pre-processing method for agent communication. In \cite{wang2020learn}, each vehicle in the vehicle networking system acts as an agent, transmitting its local CSI to the base station, which is the central controller. The central controller integrates the messages and sends local agents the vehicle-to-vehicle (V2V) link channel allocation policies to maximize the system's throughput. To avoid the significant signaling overhead that may arise from the instantaneous global CSI collected \cite{li2022applications}, each local agent learns to compress the CSI with an independent DNN.

The authors in \cite{gundougan2020distributed} design a MADRL-based method to maximize vehicle networks' packet reception rate (PRR) in cellular-free scenarios. Each vehicle agent independently selects spectrum resources based on its state and messages exchanged with other agents through V2V links.

Different for \cite{gundougan2020distributed}, \cite{xiang2022multi} proposes a decentralized spectrum access algorithm for cellular vehicle-to-everything (C-V2X) networks to maximize the total throughput of vehicle-to-infrastructure (V2I) users. Agents in \cite{xiang2022multi} learn communication through dedicated channels using a DNN-based message generator module (MGM) independent of action selection.

\cite{zhi2022deep} proposed a joint optimization scheme for mode selection and channel allocation in Device-to-Device (D2D) Heterogeneous Cellular Networks based on MADRL for millimeter wave and cellular frequency bands. The authors studied the interference problem caused by spectrum sharing between users of different cellular networks. To ensure the quality of service (QoS) of each D2D user, the agents exchange their satisfaction state information on QoS constraints to maximize user satisfaction. 

The MADRL system in \cite{pei2022intelligent} implemented a two-level unlicensed spectrum access framework consisting of a feedback cycle and an execution cycle. The base stations, as agents, transmit data to unlicensed spectrums to alleviate the pressure on cellular networks. The MADRL system is designed as a multi-agent game model, where agents share their state and reward information through broadcast to achieve the system's Nash equilibrium and maximize the network's total throughput.

\subsection{Transmit Power Control}
In wireless networks, network entities typically need to control transmission power to reduce interference with other network entities. In the MADRL system, agents can adaptively select transmit power based on the observation of the environmental changes, which is considered an effective tool for solving power control problems. Moreover, through communication, agents can obtain the interference they cause to other agents or the power allocation of other agents to find a balance between improving transmit power and reducing interference.

 \cite{khan2020centralized} proposed a power adaptive allocation algorithm based on reinforcement learning in multi-user cellular networks. The agent is the base station (BS), and the state of each BS is the local CSI and the power allocation of the previous time step. BSs exchange their power allocation information to improve the network throughput. 

\cite{zhao2020reinforcement} designs a power control algorithm for a cellular vehicle network based on MADRL. The system consists of a BS and multiple vehicle user equipment (VUEs) covered by that BS, where the agent is active V2V link. Agents reduce channel interference by sharing their channel selection from the previous time step with neighbors, maximizing the total capacity of the V2I link.

The authors of \cite{naderializadeh2021resource} proposed a radio resource management algorithm based on MADRL, where each Access Point (AP) is an agent, and each AP connects multiple user equipment devices (UEs). The data is adaptively transmitted to the associated UEs based on SNR and weight. APs exchange observations with neighboring agents, which are the weights and SNR of UEs. In addition to direct communication between APs, agents also receive feedback reports from UEs to understand the information of other APs, which has an unavoidable delay. 

In multi-user downlink small cell networks, traditional cooperative resource allocation (RA) requires collecting global CSI to calculate SNR, which is difficult to achieve in network environments. To address the difficulty of collecting global CSI in practical networks with limited direct link capacity, \cite{jang2020deep} proposes a power control algorithm based on MADRL for small cell clusters. Small cell BSs as agents only use local CSI at the transmitters and exchange in-cell sum rate through direct links to maximize the system sum rate.

To overcome the cross-layer signal interference problem between small and macro cells, the authors of \cite{xu2023distributed} designed a penalty-based Q-learning algorithm. By introducing regularization terms into the loss function, agents are encouraged to choose experiential actions with high global rewards to promote cooperation between agents. Agents share local observations through wired and wireless backhaul links to achieve balanced power allocation policies.

In \cite{nasir2019multi}, the authors introduce an interference sorting technique to handle interference information in dynamic power allocation in wireless networks. In HetNet with multiple APs and users, the interference sources of the transmitter are classified based on the received power of the corresponding interference sources at the receiver. Each transmitter (agent) not only collects CSI and QoS information from neighboring agents but also exchanges status information, such as transmit power interference through communication, and adjusts its transmit power accordingly.

\cite{gu2020deep} implementing collaborative power control and resource management in a multi-user downlink small cell network using MADRL. The optimal power allocation strategy for joint sub-carriers in IoT systems is learned through a dual deep Q network algorithm without complete instantaneous CSI. The adjacent agents share information regarding spectral efficiency, channel gain, and received power.

\subsection{Network Security}
Network security is a technology that protects network entities and information from malicious attacks. Common challenges include jamming attacks, distributed denial of service (DDoS) attacks, etc. The jamming attack refers to attackers interfering with legitimate communication channels by sending interference signals. DDoS is a distributed cyber attack aimed at depleting the network resources of the target system. Recently, the joint network attack defense method based on MADRL for multiple network entities has been extensively studied.

\cite{yao2019collaborative} uses a multi-agent Q-learning algorithm to learn distributed anti-jamming strategies for each agent. A jammer in the system initiates jamming attacks on one of the channels each time. Meanwhile, agents choose the same channel, causing co-channel interference. The agent is the legitimate user, each agent exchanges Q-values and sums them up to select the joint action that can maximize the sum of Q-values. 

The authors in \cite{xu2018interference} propose a collaborative anti-jamming algorithm based on multi-agent Q-learning in UAV communication networks. As agents, the UAV group users are usually in a competitive relationship without communication. When the agent perceives that the energy of the co-channel interference signal exceeds the threshold, the agent determines that they have been affected by the co-channel jamming, and the agents in the system switch to cooperative mode. In the cooperative mode, agents share their Q table to output joint action to counteract jamming attacks and maximize the user utility.

The authors in \cite{xia2019new} use a centralized MADRL system based on a hierarchical communication mechanism to defend against DDoS attacks. Each router, as a local agent, sends its traffic reading to the central router, which then decides the throttling rate for each router. To reduce the huge communication costs caused by the frequent exchange of information with the central agent, each local agent adds a deep deterministic policy gradient network to determine whether to send local information to the central agent.

\section{Communication Schemes}
The communication schemes refer to the learning and execution schemes of the MADRL system, which can be classified into fully centralized learning and execution, centralized training and distributed execution, and fully distributed execution. Different learning and execution schemes will be selected according to the needs of different application scenarios. 

The summary is shown in Table III.

\begin{table}
\caption*{TABLE III}
\caption*{\scshape The Category of Communication Schemes}
\label{table3}
\begin{tabular}{|l|p{6em}|p{5em}|p{4em}|p{4em}|}
\hline
\textbf{Types}  & \textbf{\makecell[c]{Traffic \\ Engineering}} & \textbf{\makecell[c]{Network \\ Access}} & \textbf{\makecell[c]{Power \\ Control}} & \textbf{\makecell[c]{Network \\ Security}} \\ \hline

\makecell[l]{Fully \\ Centralized}   & \makecell[c]{-} & \cite{wang2020learn} \cite{shamsoshoara2020autonomous} & \makecell[c]{-} & \cite{xia2019new} \\ \hline

\makecell[l]{CTDE} & \cite{you2020toward} \cite{zhang2021sac} \cite{jiang2018graph} &  \cite{gundougan2020distributed}  \cite{xiang2022multi} & \cite{khan2020centralized} \cite{zhao2020reinforcement}  & \cite{yao2019collaborative} \cite{xu2018interference} \\ \hline

\makecell[l]{Fully \\ Distributed} & \cite{zhao2020improving} \cite{kaviani2021robust} \cite{li2020routing} \cite{li2020modified} \cite{chen2021universal}  \cite{alliche2022impact} \cite{modi2023multi} \cite{qiu2021qlgr}  &  \cite{zhi2022deep} \cite{dong2022dynamic} \cite{pei2022intelligent} &  \cite{nasir2019multi}  \cite{naderializadeh2021resource} \cite{gu2020deep} \cite{jang2020deep} \cite{xu2023distributed} & \makecell[c]{-}  \\ \hline
\end{tabular} 
\end{table}

\subsection{Fully Centralized}
In the fully centralized scheme, local agents report their observations to a central agent, which decides what local agents should execute. The central agent can effectively output the globally optimal policy by utilizing instantaneous global state information. 

For example, the traditional dynamic spectrum access algorithm usually relies on global CSI \cite{wu2018high,kuang2018energy}, while the MADRL dynamic spectrum access algorithm, such as \cite{wang2020learn}, evolved from traditional methods, adopts a fully centralized structure to enable the central agent to achieve adaptive channel adjustment based on global CSI received from the local agents.

By collecting global instantaneous state information, a fully centralized model can effectively overcome non-stationary problems. However, frequent interaction between all local agents and the central agent may lead to high communication costs. In addition, communication delays may result in the central control being unable to collect complete instantaneous global states, thereby affecting agent decision-making.

\subsection{Centralized Training and Distributed Execution}
Centralized Training and Distributed Execution (CTDE) allows agents to use non-immediate global information for centralized learning, and make their own decisions independently during the execution phase. Compared with the fully centralized system, the MADRL system with the CTDE framework usually has less communication cost during execution and has been widely used in network management. Additionally, compared with distributed learning, agents learning a shared policy can reduce training parameters and accelerate convergence speed \cite{gupta2017cooperative}. Therefore, the CTDE framework has been widely used in network management. However, centralized training may lead to poor scalability. Changes in network topology may require all agents to be retrained in a centralized training system.

\subsection{Fully Distributed}
In the fully distributed scheme, each agent is trained with an independent network. Similar to CTDE, agents can overcome non-stationarity during the execution phase by exchanging messages with other agents. Compared to centrally trained models, distributed training models have better flexibility and scalability, thereby have been used in networks built on mobile network entities, such as \cite{kaviani2021robust} \cite{li2020routing} \cite{li2020modified} \cite{modi2023multi} \cite{qiu2021qlgr} \cite{nasir2019multi}, or heterogeneous networks composed of different network entities, such as \cite{zhi2022deep} \cite{xu2023distributed}.

The fully distributed learning and execution schemes align with the trend of future network management toward distribution and decentralization. However, each agent must be trained through an independent network, which can result in higher training costs.


\section{Communication Content}

The communication content refers to what information is encoded in the communication messages. The communication messages may generally be the agent's state, action, reward, or strategy. This information may be used to assist agents in completing their perception of environmental changes, learning policies from other agents, or promoting collaboration between agents. Agents in different systems interact with various types of information based on diverse application scenarios. The summary is shown in Table IV.

\begin{table}
\caption*{TABLE IV}
\caption*{\scshape The Category of Communication Content}
\label{table4}
\begin{tabular}{|l|p{6em}|p{4em}|p{6em}|p{4em}|}
\hline
\textbf{Types}  & \textbf{\makecell[c]{Traffic \\ Engineering}} & \textbf{\makecell[c]{Network \\ Access}} & \textbf{\makecell[c]{Power \\ Control}} & \textbf{\makecell[c]{Network \\ Security}} \\ \hline

State&\cite{zhao2020improving} \cite{you2020toward} \cite{li2020routing} \cite{li2020modified}  \cite{zhang2021sac} \cite{jiang2018graph} & 
\cite{gundougan2020distributed} \cite{zhi2022deep} \cite{xiang2022multi}& 
\cite{khan2020centralized} \cite{nasir2019multi} \cite{zhao2020reinforcement} \cite{naderializadeh2021resource} \cite{gu2020deep} \cite{jang2020deep} & \makecell[c]{-}
\\ \hline

Reward& \cite{chen2021universal} \cite{kaviani2021robust} \cite{modi2023multi} \cite{qiu2021qlgr}  & 
\makecell[c]{-} & 
\cite{xu2023distributed} & \cite{yao2019collaborative} \cite{xu2018interference} \\ \hline

\makecell[l]{Mixed \& \\ Other} & \cite{alliche2022impact} & \cite{wang2020learn} \cite{shamsoshoara2020autonomous} \cite{dong2022dynamic} \cite{pei2022intelligent} & \makecell[c]{-} & \cite{xia2019new} \\ \hline

\end{tabular} 
\end{table}

\subsection{State or observation}
Agents can compensate for their partial knowledge of the environment by exchanging observation (partial state) or complete state information, and its specific parameters are usually closely related to the application scenario and optimization goal of the MADRL system. 

For example, in research on traffic engineering, agents learn to select the next hop for data packets or flows, in order to improve system throughput, reduce transmission delay, and avoid congestion. Therefore, to minimize the average transmission time of each data packet, in \cite{you2020toward} and \cite{zhang2021sac}, agents exchange queue length with neighbors.

In \cite{zhao2020improving}, the objective of the research is to enhance the average throughput performance of the system that comprises ASs acting as agents. The agents collaborate by sharing their observations including the current flow in the agent, the maximum number of flows, the number of neighboring agents, and the throughput of the flow. 

Additionally, traffic engineering in wireless communication would consider the link stability or data validation. For example, the research on unmanned space self-organizing network \cite{li2020modified} shares the effective data payload with the next hop agent. In research on underwater wireless sensor networks \cite{li2020routing}, communication messages are designed to reflect agent energy and link stability to reduce sensor energy use and extend the network life cycle. 

\cite{jiang2018graph} propose Graph Convolutional Reinforcement Learning in a graph representation multi-agent system to achieve adaptive routing. The agent is a data packet, and the characteristics of the nodes on the graph serve as the state information of the agent, including the packet's current location, destination, data size, link load, and the number of adjacent data packets. Agents use GNN to learn the encoding of state messages and share them with neighboring agents.

Research on transmit power control in wireless communication networks usually aims to improve the transmission rate while reducing the impact of interference between network entities on the system. At the same time, adjusting the power allocation ratio improves the overall network throughput or reduces energy consumption. Therefore, the state information interacted by agents in related research includes power allocation ratio \cite{khan2020centralized}, interference from other agents \cite{nasir2019multi}, SNR \cite{naderializadeh2021resource}, channel gain \cite{nasir2019multi,gu2020deep}, received power \cite{gu2020deep}, and CSI \cite{xu2023distributed}.

The network spectrum access method needs to avoid excessive competition for spectrum resources among network entities and choose a balanced spectrum resource allocation policy. Therefore, agents inform other agents of channel occupancy information \cite{zhi2022deep} or interference \cite{xiang2022multi} through communication. Moreover, the MADRL system in \cite{zhi2022deep} aims to maximize user satisfaction, thereby, the agents also exchange their satisfaction with QoS constraints with each other. 

Besides, in \cite{xiang2022multi}, the agent does not directly transmit observation information but encodes the observation through a message communication network that is opposed to the action decision network. Message communication network of agents in \cite{xiang2022multi} applying a Discretization/Regularization Unit (DRU) to regularize the output during the training phase and discretize it during the execution phase so that the parameters of the communication network can be updated using gradient backpropagation.

In addition, the location information of mobile devices as agents is also considered as shared status information in the wireless communication MADRL system, such as \cite{gundougan2020distributed}.

\subsection{Reward}
The reward function and its parameters intuitively reflect the impact of the agent's decisions on the environment. Agents can better perceive their impact on the environment by exchanging rewards. For example, \cite{chen2021universal} models the CLS system as a flow routing problem to achieve the minor consumption of system joint resources. The reward function used for sharing considers the number of flows in the current agent, the number of agents that each flow needs to pass through, and the number of virtual links connected to the agent. Similar work include \cite{modi2023multi} and \cite{qiu2021qlgr}.

\cite{xu2023distributed} investigates maximizing the network overall rate by using power MADRL power control, with each cell's in-cell sum rate (ICSR) as the agent's reward. The simulation results indicate that sharing ICSR between agents can enhance cooperation and improve the overall rate. In addition, in wireless communication, agents can confirm the arrival of packets through feedback ACK messages, which is part of the reward function, such as in \cite{kaviani2021robust}.

In Q-learning, the Q-value is the maximum reward that an agent can obtain in a particular state and specific action \cite{riedmiller2005neural}. Agents can achieve the global optimal joint action by sharing Q-values. In scenarios with jamming attacks, in \cite{yao2019collaborative}, each legitimate user, regarded as an agent, learns the best joint co-channel anti-interference strategy by sharing Q-values. Similarly, in the research anti-jamming attacks, the agents in \cite{xu2018interference} share Q-tables instead of Q-values. Simulation results depict that by sharing Q-tables, agents can avoid co-channel jamming effectively, but the Q-table is about 1024 bytes, while each Q-value is only 1 bytes. Thus, agents only communicate with each other when the system is subjected to external co-channel jamming in \cite{xu2018interference}. 

\subsection{Mixed \& Other Parameters}
The communication messages between agents may contain various parameters, not just sharing one of the states, actions, or rewards. 

In the fully centralized MADRL model, such as \cite{wang2020learn}, \cite{shamsoshoara2020autonomous}, and \cite{xia2019new}, local agents submit observations to the central agent and receive action decisions information. Furthermore, \cite{wang2020learn} uses DNN to compress the observations of each local agent in the vehicular network system and further enhances it through a quantization layer to reduce network signaling costs.

The authors of \cite{alliche2022impact} design two communication modes for agents, model sharing and value sharing. In model sharing, agents would share a copy of their machine learning model parameters, and agents would share their observations during value sharing mode. The system performance using model sharing is superior to value sharing in all aspects, but the size of each value sharing information is 8 Bytes, while the size of each target model update packet is 512 Bytes.

The authors in \cite{dong2022dynamic} compared the effect of agents transmitting states, actions, and rewards concurrently in communication, as well as only exchanging partial information, on the overall spectrum utilization of the system. According to the simulation results, the system exhibited better performance with an increase in the number of types of communication parameters.


\section{Communication Object}

The communication object refers to whom the agents in the system determine to send messages to. Depending on the range of communication objects that can be reached, they are categorized as neighbor agents, all other agents in the system, or a central agent.

The summary is shown in Table V.

\begin{table}
\caption*{TABLE V}
\caption*{\scshape The Category of Communication Object}
\label{table5}

\begin{tabular}{|l|p{6em}|p{4em}|p{6em}|p{4em}|}
\hline
\textbf{Types}  & \textbf{\makecell[c]{Traffic \\ Engineering}} & \textbf{\makecell[c]{Network \\ Access}} & \textbf{\makecell[c]{Power \\ Control}} & \textbf{\makecell[c]{Network \\ Security}} \\ \hline

Neighbor&\cite{zhao2020improving} \cite{you2020toward} \cite{chen2021universal} \cite{alliche2022impact} \cite{zhang2021sac} \cite{jiang2018graph} & \cite{zhi2022deep} \cite{xiang2022multi} & \cite{nasir2019multi} \cite{zhao2020reinforcement} \cite{naderializadeh2021resource} \cite{gu2020deep} \cite{xu2023distributed} & \makecell[c]{-} \\ \hline

All&\cite{kaviani2021robust} \cite{li2020routing} \cite{li2020modified} \cite{modi2023multi} \cite{qiu2021qlgr} & \cite{gundougan2020distributed} \cite{dong2022dynamic} \cite{pei2022intelligent} & \cite{khan2020centralized}  \cite{jang2020deep} & \cite{yao2019collaborative} \cite{xu2018interference} \\ \hline

Central& \makecell[c]{-} & \cite{wang2020learn} \cite{shamsoshoara2020autonomous} & \makecell[c]{-} & \cite{xia2019new} \\ \hline

\end{tabular} 
\end{table}

\subsection{Neighbor Agents}
In a wired network, network entities such as routers or AS typically have established upstream and downstream relationships in the network topology. In this environment with relatively fixed network topology, agents usually choose to share information with neighboring entities, considering their greater impact than other agents in the network, such as \cite{zhao2020improving,you2020toward,chen2021universal,alliche2022impact,zhang2021sac,jiang2018graph}.

In addition, in the wireless network, to ensure communication quality, agents are set to communicate with other agents that can establish stable communication channels within a specific range, such as \cite{zhi2022deep,nasir2019multi,zhao2020reinforcement,naderializadeh2021resource,gu2020deep,xu2023distributed,xiang2022multi}.

\subsection{All Agents}
Network entities can naturally communicate with each other on the Internet, and agents applied on the Internet can communicate with all other agents directly, such as in \cite{modi2023multi}, \cite{yao2019collaborative} and \cite{xu2018interference}. 

In addition, in wireless communication environments, agents, such as UAVs or vehicles, broadcast their information to all other agents in the system due to device mobility or to simplify the model. For example, \cite{kaviani2021robust,li2020routing,li2020modified,qiu2021qlgr,gundougan2020distributed,dong2022dynamic,pei2022intelligent,khan2020centralized,jang2020deep}.

Besides, through deep learning or gating mechanisms, agents can selectively transmit messages to all other agents within the system, such as \cite{singh2018learning,hu2020event,sun2023learning}, to reduce communication costs.

\subsection{Central Agent}
In the fully centralized training and execution system, all agents only communicate with a central agent. The agents transmit their local information to the central agent and receive decision information from the center without communication with other local agents \cite{wang2020learn,shamsoshoara2020autonomous,xia2019new}. The central agent maintains real-time communication with all local agents, inevitably resulting in high communication costs. Therefore, in order to reduce communication costs, each agent in \cite{xia2019new} adds a Deep Deterministic Policy Gradient network to decide whether to send messages to the central agent. The central agent aggregates traffic information and makes joint action decisions that are allocated to several specific agents for execution.

Additionally, in order to enable each agent to observe the global environment in a distributed execution system, the agent communication algorithm using a central communication agent has been proposed \cite{niu2021multi,jiang2018learning,mao2020learning,sheng2022learning}. The central communication agent aggregates and encodes other agents' messages as feedback, without outputting any action decisions. However, to our knowledge, this algorithm has not yet been applied to network management instances.


\section{Communication Message Processing}
Communication message processing refers to the process of an agent integrating messages that are received from other agents. Deep Neural Networks typically require a fixed-dimension state representation as input for iteration. Therefore, incoming messages from other agents must be integrated to match input dimensions. Several methods for aggregating messages include concatenation, averaging, attention mechanisms, and neural networks. 

In addition, messages may be utilized as parameters for the reward function rather than inputs for neural networks. For example, in the wireless network, exchanging messages is to confirm whether the action is completed, such as in \cite{kaviani2021robust,li2020routing,li2020modified}. Moreover, the message may be used as the parameters for calculating the reward function \cite{modi2023multi,qiu2021qlgr,nasir2019multi}. Additionally, in the centralized system, the local agents receive action decisions from the central agent, which information would not need to be integrated \cite{wang2020learn, shamsoshoara2020autonomous,xia2019new}.

In this survey, we only focus on the message integration methods in agent communication and classify recent works into three categories, as shown in Table VI.

\begin{table}
\caption*{TABLE VI}
\caption*{\scshape The Category of Communication Message Processing}
\label{table6}
\begin{tabular}{|l|p{6em}|p{4em}|p{4em}|p{4em}|}
\hline
\textbf{Types}  & \textbf{\makecell[c]{Traffic \\ Engineering}} & \textbf{\makecell[c]{Network \\ Access}} & \textbf{\makecell[c]{Power \\ Control}} & \textbf{\makecell[c]{Network \\ Security}} \\ \hline

Concatenation&\cite{zhao2020improving} \cite{you2020toward} \cite{zhang2021sac} & \cite{zhi2022deep} \cite{dong2022dynamic} & \cite{khan2020centralized} \cite{naderializadeh2021resource} \cite{gu2020deep} \cite{jang2020deep} \cite{xu2023distributed} & \cite{yao2019collaborative} \cite{xu2018interference} \\ \hline

Average& \makecell[c]{-} & \cite{pei2022intelligent} & \makecell[c]{-} & \makecell[c]{-} \\ \hline

\makecell[l]{Neural \\ Network}&\cite{alliche2022impact} \cite{jiang2018graph}& \cite{wang2020learn} \cite{xiang2022multi} & \makecell[c]{-} & \makecell[c]{-} \\ \hline

\end{tabular} 
\end{table}

\subsection{Concatenation}
Concatenation is a common message integration method, ensuring that messages are not lost during the integration process. However, the concatenation method does not consider the message weights of different agents, and it reduces the system's scalability. The recent work that has used the concatenation method is presented in Table IV.

\subsection{Average}
The average calculation is a simple information integration method for agents' communication, such as in \cite{pei2022intelligent}. It reduces system complexity, but it is obvious that this approach overlooks the impact of the number of agents and their interdependence. Thus, it has not been widely used.

\subsection{Neural Network}
Simple neural networks or deep neural networks are used to learn to integrate messages from other agents, such as \cite{alliche2022impact,wang2020learn,xiang2022multi}.

Moreover, the agents in \cite{jiang2018graph} measure the weight of information with attention mechanism and use GCN to integrate the feature vectors in neighbor messages. Then, generate latent feature vectors as the interactive messages.


\section{Communication Constraints}
In a natural environment, various constraints, such as noise or limited bandwidth, can affect the communication performance between agents. However, currently, there is still relatively little research on how to address communication constraints between agents.

\subsection{Limited Bandwidth}
Bandwidth is frequently restricted in the natural surroundings. When the available bandwidth is being fully utilized, the messages sent by the agents may not be transmitted on time, leading to a delay in the agents' decision-making process. The network management literature studies the effect of limited system bandwidth on communication between multiple agents, including \cite{you2020toward} and \cite{xia2019new}. The experiment proves that the two works have better system performance in a limited Bandwidth environment compared to their baseline.

\subsection{Noise}
In communication, the influence of noise is inevitable. The network management literature which has considered the effects of noise on communication between agents includes \cite{li2020modified,chen2021universal,gundougan2020distributed,wang2020learn,xiang2022multi}. These works have been experimentally proven to have good robustness in systems under the influence of noise, but no denoising method has been proposed or applied.


\section{Open Issues and Future Research Direction}
This section discusses the issues and challenges of multi-agent communication in current network management. And briefly review future research directions.

\subsection{Denoise}
Noise can corrupt data and reduce the reliability of communication between agents \cite{foerster2016learning}. However, existing research on multi-agent systems in network management has not given much attention to the impact of communication noise between agents. Moreover, there is a lack of research on denoising in multi-agent communication processes. Hence, the problem of noise and interference in communication among agents is a pressing issue that requires immediate attention and resolution. 

Recent research has continuously explored the application of denoising with deep learning. Deep learning, especially CNN-based image-denoising methods, has been widely studied. Since many MARL environments can be characterized as graphs, the impact of noise on agent communication might also be optimized using image-denoising algorithms in the future.

\subsection{Scalability}
In scenarios where the number of agents varies over time, the dimension of the system state representation changes dynamically.
However, Neural networks typically require fixed-dimension state representation, posing scalability issues for MADRL systems. Many existing studies use mean calculation, attention, or deep neural networks to aggregate messages and reduce them to a fixed size. However, this approach may not be applicable in all scenarios. When new agents are added to a system, it can increase the number of messages and require changes to the action space of other agents, which can pose a significant challenge to the system's scalability.

\subsection{Synchronization and Information Delay}
In distributed or decentralized multi-agent networks, ensuring synchronization of agent states during training is a common challenge. Agents in a system rely on communication with each other to make decisions and cooperate effectively. However, if messages are delayed or agents are out of sync, it can lead to decision delays and decreased overall system performance. This is because agents may use outdated information for training or make decisions. Therefore, it is important to ensure that agents can communicate effectively and in a timely manner to optimize system performance.

\subsection{Limited Bandwidth}
In the natural environment, bandwidth cannot be unlimited. Excessive messages from agents can result in delays in transmitting information, which further reduces the system's performance. It is important to ensure that messages are transmitted efficiently in limited bandwidth. Emergent communication research seems to be an effective solution by allowing agents to self-learn to determine communication objects, content, or timing. However, the use of emergent communication in managing multi-agent networks is still relatively limited.

\subsection{Readability of Emergent Communication}
The messages of communication can emerge from agents through deep learning, such as in \cite{jiang2018graph,wang2020learn,xiang2022multi}, but the emergent messages may be unreadable to humans. This makes it difficult to evaluate communication effectiveness and formulate relevant specifications for industrial applications. Therefore, emergent message interpretation is one of the future research directions of agent communication.

\subsection{Communication Security}
In the application scenario of network management, agents may contain privacy-sensitive information, such as user location, traffic, device power, etc., through sharing state or observation. If the agent containing privacy-sensitive information is left unprocessed, hackers may gain access to the entire system by attacking and controlling some agents. In this case, the information security issues in agent communication are worth studying.


\section{Conclusion}
In conclusion, we conducted a comprehensive survey of the application of MADRL in network management, especially a detailed analysis of research involving communication between multiple agents. First, we introduced the research contributions in current network management that involve communication between numerous agents. Then, we compared and classified agent communication work in detail from the aspects of learning and training schemes, communication objects, communication content, and communication constraints. Finally, we analyzed the challenges of Multi-agent communication applied in current network management and the future research directions.

\section*{Acknowledgment}

This research is supported in part by the Huawei Technologies Co., Ltd., in part by the National Natural Science Foundation of China (Grant No. 92067109, 61873119, 62211530106), and in part by Shenzhen Science and Technology Program (Grant No. ZDSYS20210623092007023, GJHZ20210705141808024).

\hspace*{\fill}

\ifCLASSOPTIONcaptionsoff
  \newpage
\fi
\begin{refcontext}[sorting = none]
\printbibliography
\end{refcontext}
\end{document}